\documentclass[reprint,amsmath,amssymb,aps,prb,showpacs,showkeys]{revtex4-1}
\usepackage{graphicx}
\usepackage{dcolumn}
\usepackage{bm}
\usepackage{natbib}
\usepackage{color}
\usepackage[normalem]{ulem}
\usepackage{bm}              
\usepackage[T1]{fontenc}

\newcommand{\angstrom}{\mbox{\normalfont\AA
}}
\def \pst {Pb$_{1-x}$Sn$_x$Te}
\def \pss {Pb$_{1-x}$Sn$_x$Se}

\begin{document}
\bibliographystyle{plainnat}
\title{
Alloy broadening of the transition to the non-trivial topological phase 
of \pst
}

\author{ A.~{\L}usakowski}
\affiliation{Institute of Physics, Polish 
Academy of Sciences, Al. 
Lotnik\'{o}w 32/46, 02-668 Warsaw, Poland}
\author{P. Bogus{\l}awski}
\affiliation{Institute of Physics, Polish 
Academy of Sciences, Al. 
Lotnik\'{o}w 
32/46, 02-668 Warsaw, Poland}
\affiliation{Institute of Physics, 
University of Bydgoszcz, ul. 
Chodkiewicza 
30, 85-072 Bydgoszcz, Poland}
\author{T. Story}
\affiliation{Institute of Physics, Polish 
Academy of Sciences, Al. 
Lotnik\'{o}w
32/46, 02-668 Warsaw, Poland}

\begin{abstract}%

Transition between the topologically trivial and non-trivial phase of \pst\ 
alloy is driven by the increasing content $x$ of Sn, or by the hydrostatic 
pressure for $x<0.3$. We show that a sharp border between these two 
topologies exists in the Virtual Crystal Approximation only. In more 
realistic models, the Special Quasirandom Structure method and the 
supercell method (with averaging over various atomic configurations), the 
transitions are broadened. We find a surprisingly large interval of alloy 
composition, $0.3<x<0.6$, in which the energy gap is practically vanishing. 
A similar strong broadening is also obtained for transitions driven by 
hydrostatic pressure. Analysis of the band structure shows that the alloy 
broadening originates in splittings of the energy bands caused by the 
different chemical nature of Pb and Sn, and by the decreased crystal 
symmetry due to spatial disorder. Based on our results of {\em ab initio} 
and tight binding calculations for \pst\ we discuss different criteria of 
discrimination between trivial and nontrivial topology of the band structure 
of alloys. 

\end{abstract}
\maketitle

\section{Introduction}

IV-VI compounds and their substitutional alloys constitute the most 
important 
family 
of topological crystalline insulators. They comprise in 
particular PbTe and SnTe, which differ by the order of 
levels at the L point of the Brillouin Zone (BZ): in PbTe, 
the symmetry of the valence band maximum, VBM, 
(conduction band minimum, CBM) is L$_{6+}$ (L$_{6-
}$), while in the topologically non-trivial SnTe the order 
is inverted, which is referred to as the inverted band 
structure with a negative energy gap $E_{gap}$. We stress that a 
negative gap still means an insulating situation with the open gap 
$|E_{gap}|>0$. An analogous situation occurs for the 
second pair of IV-VI compounds, PbSe and 
SnSe.\cite{nimtz,khoklov} Pseudobinary alloys of those compounds 
offer a unique possibility of detailed studies of the 
transition between topologically trivial and nontrivial 
phases. Such studies were performed for the Pb$_{1-
x}$Sn$_{x}$Se alloy. In this case, Angle Resolved 
Photoemission Spectroscopy  data showed that in the 
relatively wide composition window $0.2 < x < 0.4$ the 
transition to the nontrivial phase is driven by the 
decreasing temperature. \cite{dziawa, tanaka,xu,wojek} It is believed 
that 
the observed inversion of the band gap character is 
mainly induced by the decrease of the lattice constant 
due to thermal contraction. This premise is supported both by 
the experimental pressure dependencies of the lead and 
tin chalcogenides band gaps\cite{nimtz,khoklov} and by theoretical 
analysis.\cite{barone}  

Previous theoretical studies of IV-VI alloys were mostly 
conducted within the Virtual Crystal Approximation 
(VCA), which essentially leads to almost linear 
dependencies of alloy properties on composition. VCA 
explained a number of important features observed in experiment, 
such as the existence of zero gap Dirac-like surface states 
in thick slabs.\cite{rb1, rb2,rb3} As we show, the VCA applied to 
the \pst\ alloy predicts a sharp transition between the 
topologically trivial and nontrivial phases driven by the 
increasing content of Sn, or by the applied hydrostatic 
pressure (i.e., by the decreasing lattice parameter). One 
should notice, however, that within the VCA an alloy has 
the full point and translational symmetry of the rock salt 
structure. On the other hand, in real alloys there always 
is the chemical disorder, and those symmetries are 
missing. 

Our goal is to examine to what extent the results of the 
VCA for \pst\ are realistic. 
We compare the 
VCA results with the band structures obtained within 
supercell models of \pst\ alloy which explicitly differentiate 
the two cations, Pb and Sn.
We show that the presence of two types of cations 
together with alloy randomness drastically modifies the 
band structure. In contrast to the VCA predictions, the 
dependence of $E_{gap}$ on the composition is 
non-linear. Instead of the sharp transition from the direct 
to the inverted structure, there is a wide composition 
range $0.3<x<0.5$ with the zero band gap. For the sake of simplicity,
in the following the case with the $E_{gap}$ below 3 meV
is referred to as the zero gap case. A similar 
anomaly characterizes also the pressure induced 
transition to the non-trivial phase, which can take place 
for Pb-rich \pst\ with $x<0.3$, when $E_{gap}>0$. With 
the decreasing lattice constant the energy gap becomes 
negative, and the band structure inverted. However, in 
contrast to the VCA prediction of a sharp transition, 
there is a finite range of pressures (or lattice constants) in 
which the band gap vanishes. 

Analysis of the band structure is obscured by the fact that 
there is no clear criterion which determines the sign of 
$E_{gap}$, i.e., of the (non)trivial topological character 
of the alloy. Actually, such a distinction is not obvious 
even for PbTe and SnTe (or PbSe and SnSe). Indeed, the 
criterion based on the $Z_2$ topological index, suitable  
for systems with topological protection by time reversal symmetry, 
suggests triviality of both compounds. 
This feature is due to a peculiarity of the band structure 
with the direct gaps located at four nonequivalent $L$ 
points in the BZ. To properly characterize these 
compounds, another topological index, the mirror Chern number 
(MCN), was invoked.\cite{hsieh} The MCN can be 
calculated only for systems with mirror symmetry 
planes, what is satisfied for the (110) planes of the rock-
salt structure, and in the average sense in IV-VI alloys 
\pst.\cite{fu_kane} Recently, we showed that also the 
spin Chern number\cite{prodan} (SCN) allows to 
distinguish between the topology of PbTe and 
SnTe.\cite{lusak_jasz} Contrary to the MCN, the SCN has a 
much broader range of applicability, because it can be 
calculated for crystals with the translational symmetry 
but without the point symmetry, such as random alloys 
models of \pst\ considered in the present paper. 

The anomalies found for the pressure dependence of 
$E_{gap}$ of \pst\ are reflected also in the MCN and 
SCN, which in the window of lattice constant $a$ of the vanishing 
energy gap assume almost random,  highly non-
monotonic values, and can even vanish in the cases 
which we would classify as the topologically non-trivial, 
i. e., with non-positive energy gap. 
Due to all that, and this is one of the conclusions of the 
paper, it turns out that in most cases single 
calculations for a sample with a given tin's concentration 
and the spatial distribution of cations  cannot give the 
unique answer to the question of positivity or negativity 
of the energy gap. 
A simple way, and actually the only 
one we know, to uniquely determine the sign of 
the calculated energy gap, and the topological triviality, 
is to repeat calculations for different lattice constants.

In the next Section, technical details of the calculations 
are presented. Band structure, wave functions, MCN, and 
SCN as functions of the lattice constant for pure PbTe 
are analyzed in Section III. We also extend the analysis 
to \pst\ within the VCA, and show that a sharp topological 
transition as a function of the composition $x$ takes 
place. Section IV is devoted to more realistic models of 
\pst\ mixed crystals. The topological indices are 
calculated for 8 and 16 atom supercells. We show that 
the presence of two different atomic species, Pb and Sn, 
leads to additional splittings of the energy bands and to significantly 
different behavior comparing to 
the VCA description. The dependence of $E_{gap}$ on 
the composition for larger supercells is studied using the 
Special Quasirandom Structures (SQS) method.\cite{zunger} 
Section V concludes the paper. 

%
\section{Technical details of calculations}
\subsection{Density Functional Theory calculations}  
We use the open-source OpenMX package for DFT 
calculations \cite{openmx} with fully relativistic 
pseudopotentials. The calculations are done using 
Ceperly-Alder \cite{CA} LDA exchange-correlation 
functional. For Sn, we use the original pseudopotentials 
distributed with OpenMX. The pseudopotentials for Pb 
and Te with four and six valence electrons, respectively, 
were generated using the program ADPACK distributed 
with the OpenMX package. The input parameters 
for Pb and Te were described in Ref. 
\onlinecite{lusakowski1}. 
The reason for generating new pseudopotentials was 
twofold. First, to properly study the band structure of 
alloys we must employ large supercells. The 
pseudopotentials distributed with OpenMX package were 
generated assuming 14 and 16 valence electrons for Pb 
and Te, respectively. Thus, to reduce the computational 
time we generate pseudopotentials with lower numbers 
of valence electrons.  The second reason is more 
important from the physical point of view. As it is well 
known the local density approximation underestimates 
energy gaps. This fact is particularly important for 
compounds containing heavy elements like PbTe. Due to the 
strong spin-orbit interaction for $p$(Pb) orbitals, their 
energy levels are lower than those of $5p$(Te),  what 
results in the inverted band structure. 
This problem was discussed in Ref. \onlinecite{lusakowski1}, 
and solved by a proper adjustment of the spin-orbit coupling for Pb.
Here, we use this approach.

\subsection{Calculations of the Chern 
Numbers}
The prescription for calculation of the SCN 
for 
systems where spin is not a good quantum 
number is presented in the paper by 
Prodan.\cite{prodan} He proposed to 
consider an operator 
\begin{equation}
\label{m1}
 Q({\bm k})=P({\bm k})\Sigma P({\bm k}), 
\end{equation}
where $P({\bm k})$ is the projection 
operator on the valence band states 
subspace and 
$\Sigma$ is the $z$-th component of the 
spin operator. 
The numerical procedure for calculation of 
the Chern number is 
clearly described in Ref. 
\onlinecite{suzuki}, and the method of 
obtaining both SCN and MCN is presented in 
Ref. \onlinecite{prodan}.  
However, because in our calculations the 
atomic pseudoorbitals on different lattice 
sites are not orthonormal (what is the case 
in most tight binding calculations), 
certain technical points should be 
explained.

In the tight binding approximation (TBA) 
the Bloch wave functions 
$\psi_{{\bm 
k}n}({\bm r})$ for a given wavevector ${\bm 
k}$ and the band index $n$ 
\begin{equation}
 \label{eq1}
 \psi_{{\bm k}n}({\bm 
r})=\sum_{\alpha=1}^Ma_{{\bm 
k}n\alpha}\chi_{{\bm 
k}\alpha}({\bm r}),
\end{equation}
where $a_{{\bm k}n\alpha}$ are complex 
coefficients and $M$ functions 
\begin{equation}
 \label{eq2}
 \chi_{{\bm k}\alpha}({\bm 
r})=\frac{1}{\sqrt{N}}\sum_{{\bm 
R}}e^{i{\bm k}{\bm R}}\varphi_{\alpha}({\bm 
r}-{\bm R}-{\bm 
\tau}_{\alpha})
\end{equation}
constitute the basis. In Eq.~(\ref{eq2}) 
$\alpha$ denotes the type of 
spinorbital 
($s,\ p,\ d\ ...$, spin direction), ${\bm 
R}$ numerates the positions 
of $N$ elementary cells and $\tau_{\alpha}$ 
is a vector describing the 
center of the spinorbital $\alpha$ in a 
cell ${\bm R}$. 
The functions $\chi_{{\bm k}\alpha}({\bm 
r})$ are not orthonormal:
\begin{eqnarray}
 \label{eq3}
  <\chi_{k\alpha}|\chi_{k\beta}>\equiv 
S_{\alpha\beta}({\bm k}) \nonumber 
\\
=\sum_{{\bm 
R}}e^{i{\bm k}{\bm R}}\int d{\bm r} 
 \varphi_{\alpha}({\bm r}-{\bm 
\tau}_{\alpha})
 \varphi_{\beta}({\bm r}-{\bm R}-{\bm 
\tau}_{\beta}).
\end{eqnarray}
As a consequence, the coefficients $a_{{\bm 
k}n\alpha}$ are calculated 
not from a simple eigenproblem, but from 
the generalized eigenproblem:
\begin{equation}
\label{eq4}
H_{\alpha\beta}({\bm k})a_{{\bm 
k}n\beta}=E_{{\bm 
k}n}S_{\alpha\beta}({\bm k})a_{{\bm 
k}n\beta}, 
\end{equation}
where the Hamiltonian $H_{\alpha\beta}$ is 
in the basis of functions 
$\chi_{{\bm 
k}\alpha}({\bm r})$ and the summation over 
$\beta$ is implied.

The periodic part of the Bloch wave function 
reads
\begin{equation}
\label{eq5}
u_{{\bm 
k}n}(r)=\frac{1}{\sqrt{N}}\sum_{{\bm 
R}\alpha}e^{i{\bm k}({\bm R}-{\bm 
r})}a_{{\bm k}n\alpha}\varphi_{ \alpha
}({\bm r}-{\bm R}-{\bm \tau}_{\alpha}).
\end{equation}
The method proposed in Ref. 
\onlinecite{suzuki} requires the knowledge 
of 
matrix elements $<u_{{\bm k}n}|u_{{\bm 
k_1}m}>$.
Simple calculations lead to the following 
expression
\begin{eqnarray}
\label{eq6}
 <u_{{\bm k}n}|u_{{\bm k_1}m}>=
 \sum_{\alpha\alpha_1}a^*_{{\bm 
k}n\alpha}a_{{\bm k_1}m\alpha_1}\nonumber 
\\
 \\
 \sum_{{\bm R}}\int d{\bm r} 
e^{i({\bm k}-{\bm k_1}){\bm r}+i{\bm 
k_1}{\bm R}}
 \varphi_{\alpha}({\bm r}-{\bm 
\tau}_{\alpha})
 \varphi_{\alpha_1}({\bm r}-{\bm R}-{\bm 
\tau}_{\alpha_1}).\nonumber
\end{eqnarray}
Assuming that the modulus of ${\bm 
\Delta}={\bm k}-{\bm k_1}$ is small,  
this expression may be approximated by
\begin{eqnarray}
\label{eq7}
 <u_{{\bm k}n}|u_{{\bm k}+{\bm \Delta} 
m}>\approx \nonumber \\
 \\
 \sum_{\alpha\beta}a_{{\bm k}n\alpha}^* 
\left(S_{\alpha\beta}({\bm k}+{\bm \Delta} 
)+i \Delta_i 
Z^i_{\alpha\beta}({\bm k}+{\bm 
\Delta})\right)a_{ {\bm k}+{\bm \Delta} 
m\beta}\nonumber
\end{eqnarray}
where 
\begin{equation}
\label{eq8}
 Z^i_{\alpha\beta}({\bm k})=\sum_{{\bm 
R}}e^{i{\bm k}{\bm R}}\int d{\bm 
r} 
 \varphi_{\alpha}({\bm r}-{\bm 
\tau}_{\alpha})r_i
 \varphi_{\beta}({\bm r}-{\bm R}-{\bm 
\tau}_{\beta}).
\end{equation}
The formula (\ref{eq7}) may be directly 
applied to the calculations of 
the Chern number for a chosen two 
dimensional plane in the reciprocal 
space according to the prescription 
proposed in Ref. \onlinecite{suzuki}. 
However, in the present paper we calculate 
both SCN and the MCN what 
requires additional steps. \\

Using the Cholesky factorization the 
overlap matrix $S_{\alpha\beta}({\bm 
k})$ can be expressed as the product of two 
matrices
\begin{equation}
\label{eq9}
 S_{\alpha\beta}({\bm 
k})=U^{\dagger}_{\alpha\gamma}({\bm k})
U_{\gamma\beta}({\bm k}),
\end{equation}
where the matrix $U_{\alpha\beta}({\bm k}$ 
is the upper triangular 
matrix. 
Using this matrix, for each ${\bm k}$ one 
finds a new set of basis 
functions
\begin{equation}
\label{eq10}
 \zeta_{{\bm k}\alpha}({\bm r}) = 
\chi_{{\bm 
k}\beta}({\bm r})U^{-1}_{\beta\alpha}({\bm 
k}),
\end{equation}
which are orthonormal, $<\zeta_{{\bm 
k}\alpha}|\zeta_{{\bm 
k}\beta}>=\delta_{\alpha\beta}$. 
In this basis we have to solve a simple 
eigenvalue problem
\begin{equation}
\label{eq11}
 \tilde{H}_{\alpha\beta}({\bm k})c_{{\bm 
k}n\beta} = E_n({\bm k})
c_{{\bm k}n\alpha}, 
\end{equation}
where 
\begin{equation}
\label{eq12}
\tilde{H}_{\alpha\beta}({\bm 
k})=\left(U^{\dagger}({\bm 
k})\right)^{-
1}_{\alpha\gamma}H_{\gamma\delta}({\bm 
k})U^{-1}_{\delta\beta}({\bm k})
\end{equation}
and the eigenvectors are normalized
\begin{equation}
 \label{eq13}
c^*_{{\bm k}n\alpha}c_{{\bm k}m\alpha} = 
\delta_{mn}.
\end{equation}
The following steps are based on the paper 
by Prodan.\cite{prodan} We 
focus on the SCN, and for the MCN the steps 
are analogous. 
We construct the projection operator on the 
valence band states in the 
$\zeta$ basis
\begin{equation}
 \label{eq14}
\tilde{P}_{\alpha\beta}({\bm 
k})=\sum_{n=1}^{N_v}
c_{{\bm k}n\alpha}c^*_{{\bm k}n\beta}, 
\end{equation}
where $N_v$ is the number of valence band 
states. \\
The matrix of the $z$-th component of the 
spin operator in the basis 
$\chi$ 
reads
\begin{equation}
\label{eq15}
 \Sigma_{\alpha\beta}({\bm 
k})=S_{\alpha\gamma}({\bm 
k})\Sigma^0_{\gamma\beta}, 
\end{equation}
where $\Sigma^0_{\alpha\beta}$ is the 
diagonal matrix with values $\pm 
1$, 
depending on the spin of the $\alpha$-th 
spinorbital, Eq.~(\ref{eq2}). 
After transforming the above operator to 
the $\zeta$ basis, 
$\tilde{\Sigma}$, 
we construct the operator
\begin{equation}
 \label{eq16}
\tilde{Q}_{\alpha\beta}=\tilde{P}_{\alpha\gamma}\tilde{\Sigma}_{\gamma
\delta}
\tilde{P}_{\delta\beta}. 
\end{equation}
The matrix $\tilde{Q}_{\alpha\beta}$ has 
three groups of eigenvalues. 
There are $M-N_v$ vanishing eigenvalues, 
which correspond to the 
conduction band states. 
Among the remaining $N_v$ eigenvalues, 
$N_v/2$ are positive and 
$N_v/2$ negative. 
For each ${\bm k}$ we take eigenvectors 
corresponding to positive 
(negative) eigenvalues and we calculate 
$C_{s+}$ ($C_{s-}$) SCN.

\subsection{Projection on the Anion 
Orbitals} 
The transition from the topologically 
trivial to nontrivial phase is 
closely related to the  content of anion 
$p$  orbitals  in 
the  wave functions from the VBM. In the 
TBA, the  wave functions are  build from the 
spinorbitals of the atoms constituting the 
supercell (see  equations (\ref{eq1}) and 
(\ref{eq2})). From the output of the 
calculations it is possible to draw out the 
complex coefficients $a_{{\bm  k}n\alpha}$, 
which describe the 
wave function in the TBA for a given  
wavevector  ${\bm k}$ in the 
$n$-th band. The content of anion $p$  
orbitals is defined as 
\begin{equation}  
C_{Te} = \sum_{\alpha n}|a_{{\bm 
k}n\alpha}|^2 
\end{equation}
where the sum over $\alpha$ runs over the 
anion $p$  spinorbitals. 

The wavevector ${\bm k}$ corresponds to the point in 
the BZ,  where the main energy gap is located. For 
example, in the case of the $2\times 2\times 2$ supercell 
the  main  energy gap is  located at  ${\bm k=0}$ of the 
"folded" BZ.  The sum over $n$  takes into  account the two
top  valence bands.  In an analogous way the contents of cation $p$ 
orbitals, $C_{Pb}$ and $C_{Sn}$, can be defined. 

In the case of pure PbTe where the symmetry 
of the wave functions 
of the  top  valence band is $L_{6+}$ the 
value of $C_{Te}$ is nonzero 
while for SnTe $C_{Te}$=0  because the 
corresponding wave functions 
are of $L_{6-}$ symmetry and do  not  
contain anion $p$ 
spinorbitals.  Results of numerous 
calculations  for the $2\times 2\times 
2$ supercells for \pss\ clearly indicate 
that the  analogous  results 
hold  for alloys, despite the fact that 
such 
crystals, in  general, do not have the 
cubic symmetry. Comparing the 
values of $C_{Te}$ with the  analysis of 
the energy gap as a function of 
the lattice constant we get  the  result 
that for a trivial insulator $C_{Te}$ is 
about 1.7 while for  a  nontrivial one it 
almost vanishes. 
Because we use OpenMX  program with the 
wave functions basis consisting of atomic 
non orthonormal  
pseudoorbitals, the coefficients  $a_{{\bm 
k}n\alpha}$ are not 
normalized to unity and  this is  why the 
value of $C_{Te}$ depends on 
the size of the supercell. For example,  
for a 64  atom supercell it is ~1.7, while 
for the 2 atom cell it is about 1.5.  
Our proposition is that the dependence of  
$C_{Te}$ on  the lattice constant may 
determine the topological triviality or  
nontriviality  of the given disordered 
system.  The main 
advantage of using $C_{Te}$ is that it can 
be calculated relatively quickly compared 
to  the calculations  of SCN or MCN, 
which are very time consuming for larger  
systems.\\ 

\section{Systems with the $O_h$ symmetry}
\subsection{P\lowercase{b}T\lowercase{e}}
\begin{figure}
\includegraphics[width=\linewidth]{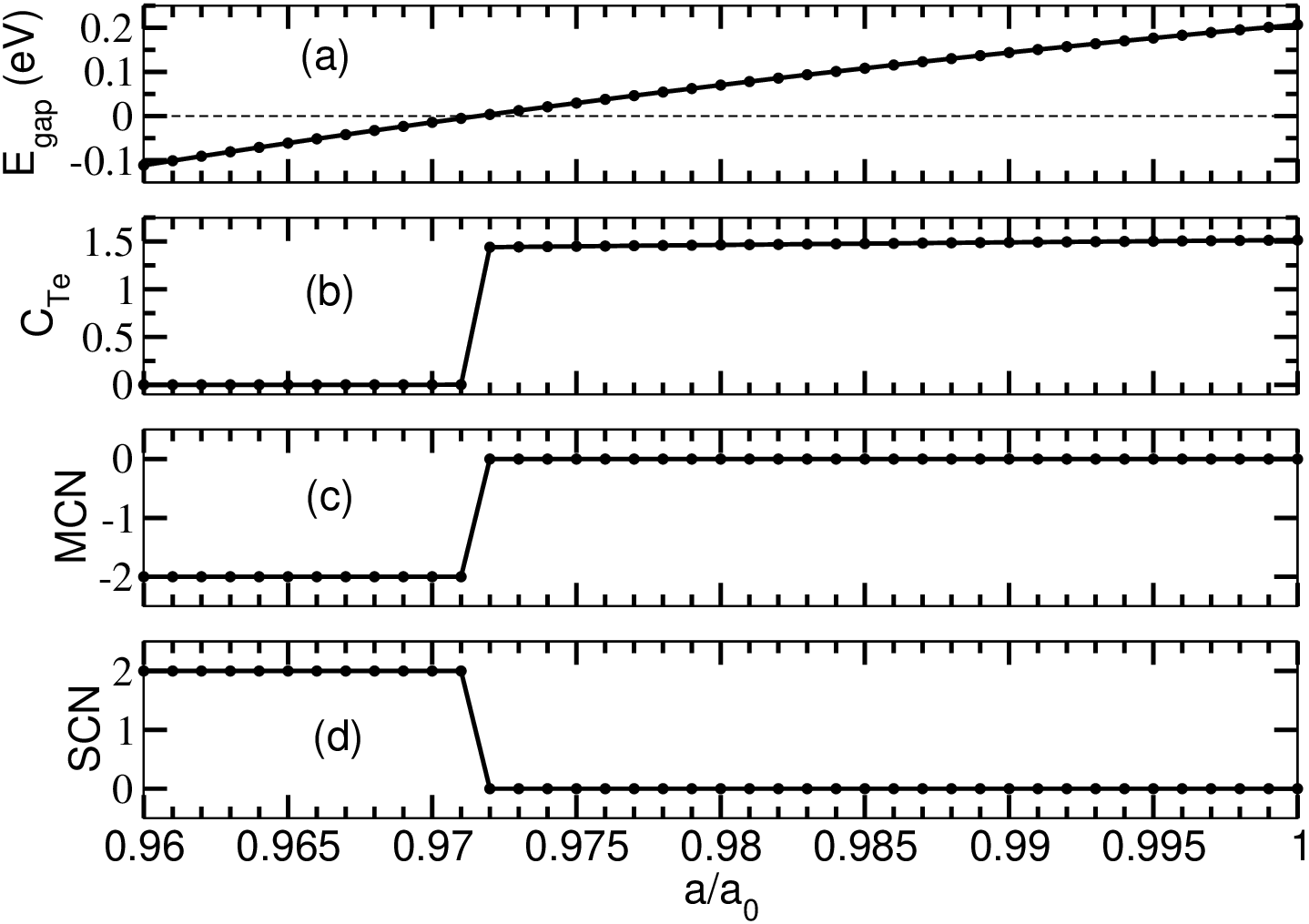}
\caption{\label{fig1}
Lattice constant dependence of (a) the PbTe 
band gap, (b) parameter 
$C_{Te}$ and topological indices (c) MCN, and (d) SCN.}
\end{figure}

As it was already mentioned, the energy gap 
of PbTe decreases with the 
decreasing lattice constant, as it is shown 
in Fig. \ref{fig1}a. The band 
gap vanishes for $a/a_0\approx 0.971$, 
where $a_0=6.46$~\AA\ is the 
equilibrium lattice constant of PbTe. With 
the further decrease of the 
lattice constant it re-opens as   
negative.

\begin{figure}
\end{figure}
\begin{figure}
\includegraphics[width=\linewidth]{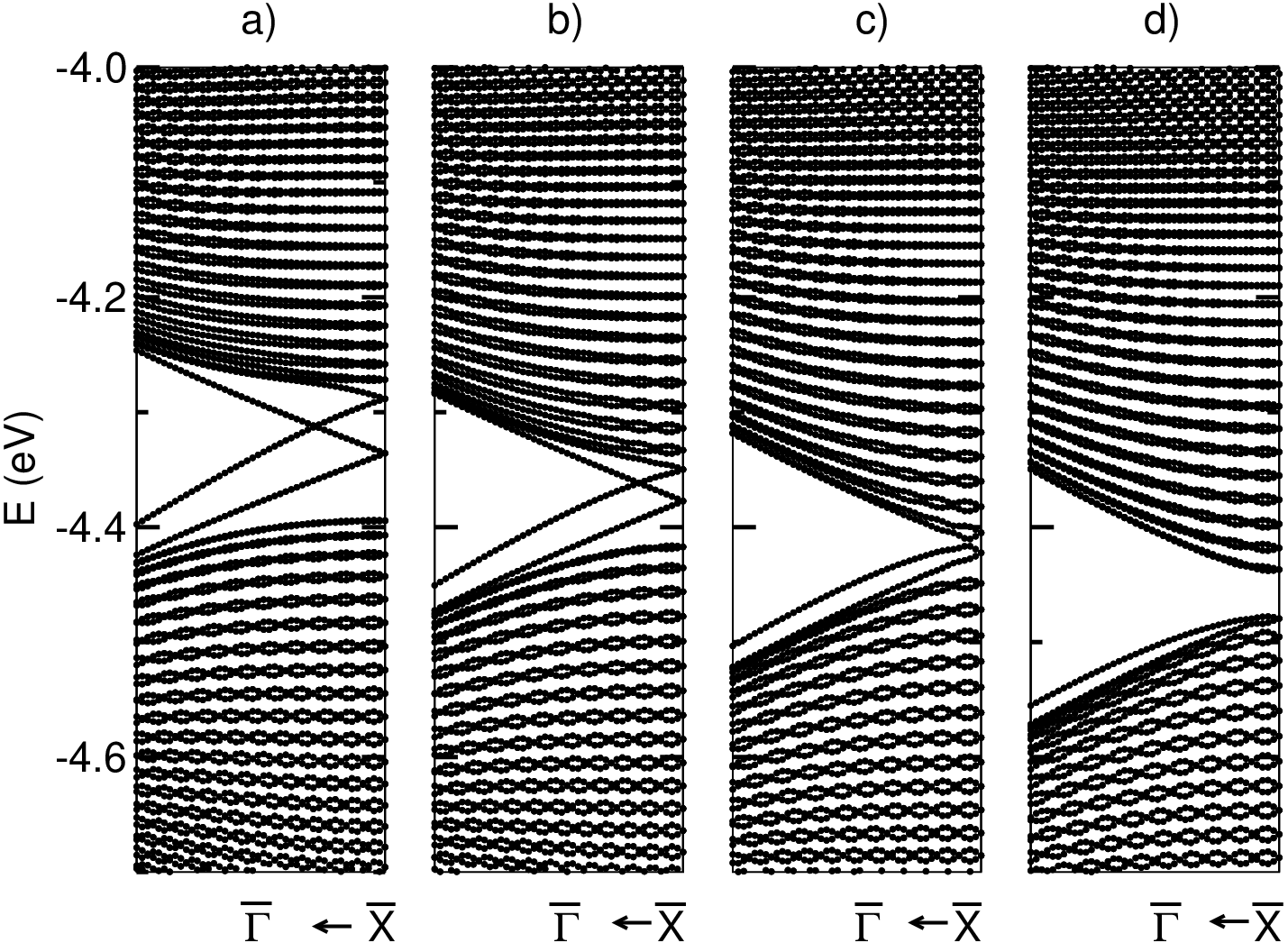}
\caption{\label{fig2}Energy levels for a 
200-atom PbTe layer oriented in 
the [001] direction for different lattice 
constant: (a) $a/a_0=0.960$, 
(b) $a/a_0=0.965$, (c) $a/a_0=0.970$ and (d) 
$a/a_0=0.975$. }
\end{figure}

\begin{figure}
\includegraphics[width=\linewidth]{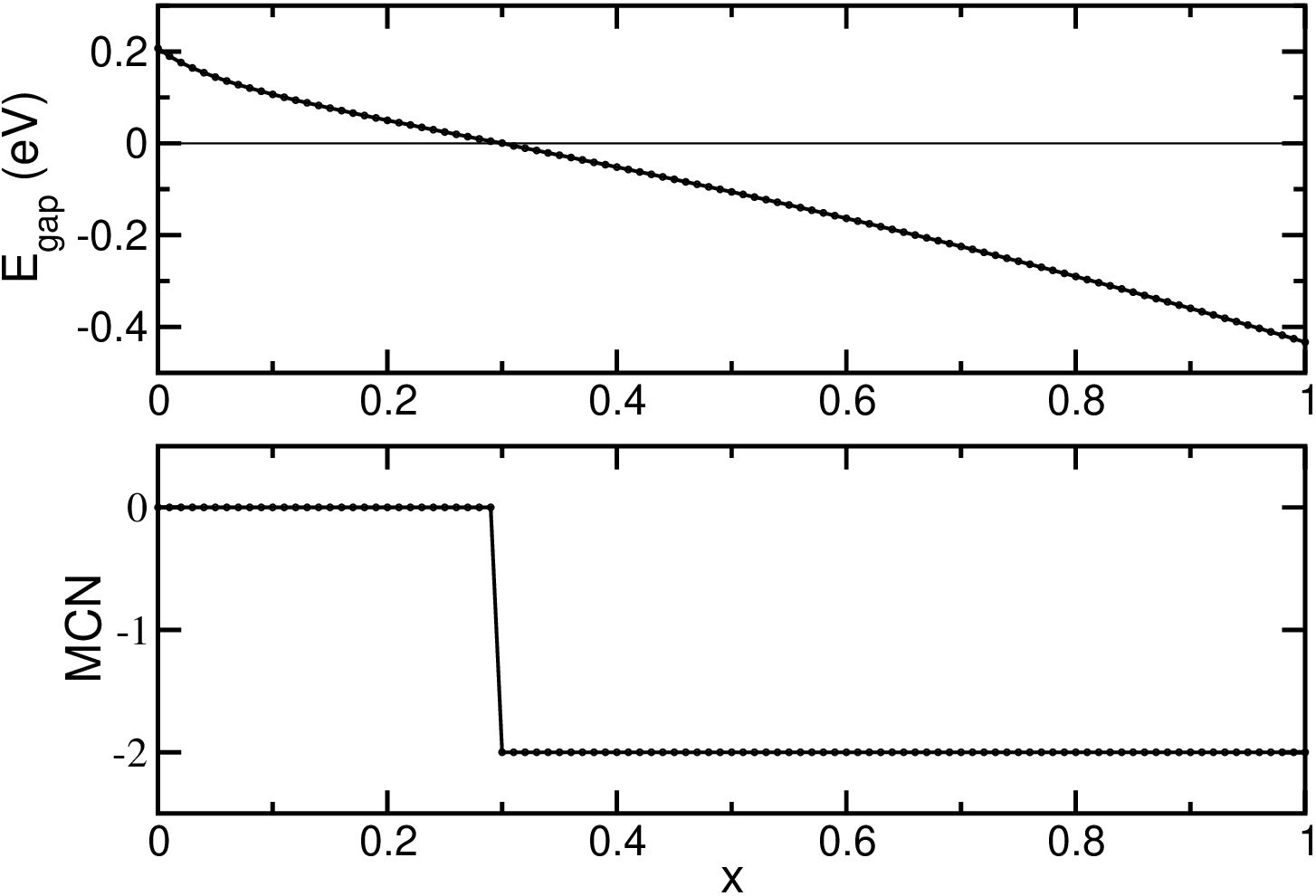}
\caption{\label{fig3} Dependence of (a) the 
energy gap at the $L$ point and (b) the  
Mirror Chern Number on the Sn content $x$ 
for \pst\ in the  VCA. }
\end{figure}

The transition from positive to negative 
energy gap coincides with the  
jump of  $C_{Te}$ from ~1.5 to approximately 
zero, 
see Fig. \ref{fig1}b. For all values of the 
lattice 
constants we calculated the MCN for the 
(110) plane, and the results are 
shown in Fig. \ref{fig1}c. The SCN, like the MCN, can be 
calculated only for two-dimensional plane 
with periodic boundary 
conditions. Let ${\bm b_1}$, 
${\bm b_2}$ and  ${\bm b_3}$ be reciprocal 
lattice vectors which span 
the 
primitive PbTe cell in  the reciprocal 
space. For a given 
$0\le z\le 1$ let us consider the points 

\begin{equation}
 {\bm p}(z)=x{\bm b_1}+y{\bm b_2}+z{\bm 
b_3}
\end{equation}
where $0\le x,y \le 1$. For this 
parallelogram we can calculate the $z$-
dependent SCN. 

The results for $z=0$ are presented in Fig. 
\ref{fig1}d. Let us notice 
that the parallelogram for $z=0$ crosses 
two $L$ points in the Brillouin 
zone. The results presented in 
Fig.~\ref{fig1} are consistent with the 
presence of surface gapless states. In 
Fig.~\ref{fig2} 
we show the energy levels along the 
$\bar{X}\rightarrow \bar{\Gamma}$ 
direction of the reduced two dimensional 
BZ of the PbTe 
layer grown along the [001] direction and 200 monolayers thick. 
We see that the bulk -- boundary 
correspondence theorem is satisfied.

\subsection{\pst\ in the Virtual Crystal 
Approximation} 

Our LDA calculations for PbTe and SnTe give two 
sets of the corresponding TBA parameters. 
Taking their composition weighted averages 
we obtain the TBA 
parameters for \pst\ in the VCA. 
For Sn, we use pseudopotentials generated 
by us with 4 valence electrons. 
In Fig.~\ref{fig3} we show both the 
$E_{gap}$ at the $L$ point and the 
MCN as the functions of composition. As in 
the case of PbTe, where a sharp 
transition between trivial and nontrivial 
topological phases takes place 
at a well defined value of the lattice 
constant, in the case of \pst\ a 
sharp transition  takes place at the well 
defined Sn content. This is in 
contrast to the more realistic models of 
\pst\ considered in the next 
Section.  

\section{Mixed 
P\lowercase{b}$_{1-
x}$S\lowercase{n}$_{x}$T\lowercase{e} 
crystals} 
\subsection{Topological properties of band 
structures}

We now turn to calculations, in which the two cations, 
Pb and Sn, are explicitly distinct. 

We start with the simplest case of the  
$1\times 1 \times 1$ supercell 
Pb$_3$Sn$_1$Te$_4$ containing eight atoms. 
As it turns out, most of  
the features characterizing larger 
supercells, 
with different numbers of  Sn atoms and 
arbitrary spatial distribution of 
cations, are observed  already  for this 
case.  

Energy gaps of PbTe are located at 
the four nonequivalent $L$ points of the BZ, 
$L = \frac{\pi}{a}(\pm 1,\pm 1,1)$ or, in the case of SnTe, near 
these points. 
The Brillouin zones of supercells are "folded" relative to that of for 
pure PbTe, and in the literature their high symmetry points are denoted 
by various symbols.

Here, we denote by $L_{gap}$ the point where for pure PbTe, the direct 
energy gap is the  smallest.  
The coordinates these $L_{gap}$ points  depend on the supercell.  For  
example for $1\times 1 \times 1$  supercell $L_{gap} = 
\frac{\pi}{a}(1,1,1)$,   for  $1\times 1 \times 2$  supercell  
considered later, $L_{gap} =   \frac{\pi}{a}(1,1,0)$, where $a$ denotes  
lattice constant of the fcc lattice.  The $\Gamma$ point always 
corresponds to ${\bm  k=0}$. 

\begin{figure}
\includegraphics[width=\linewidth]{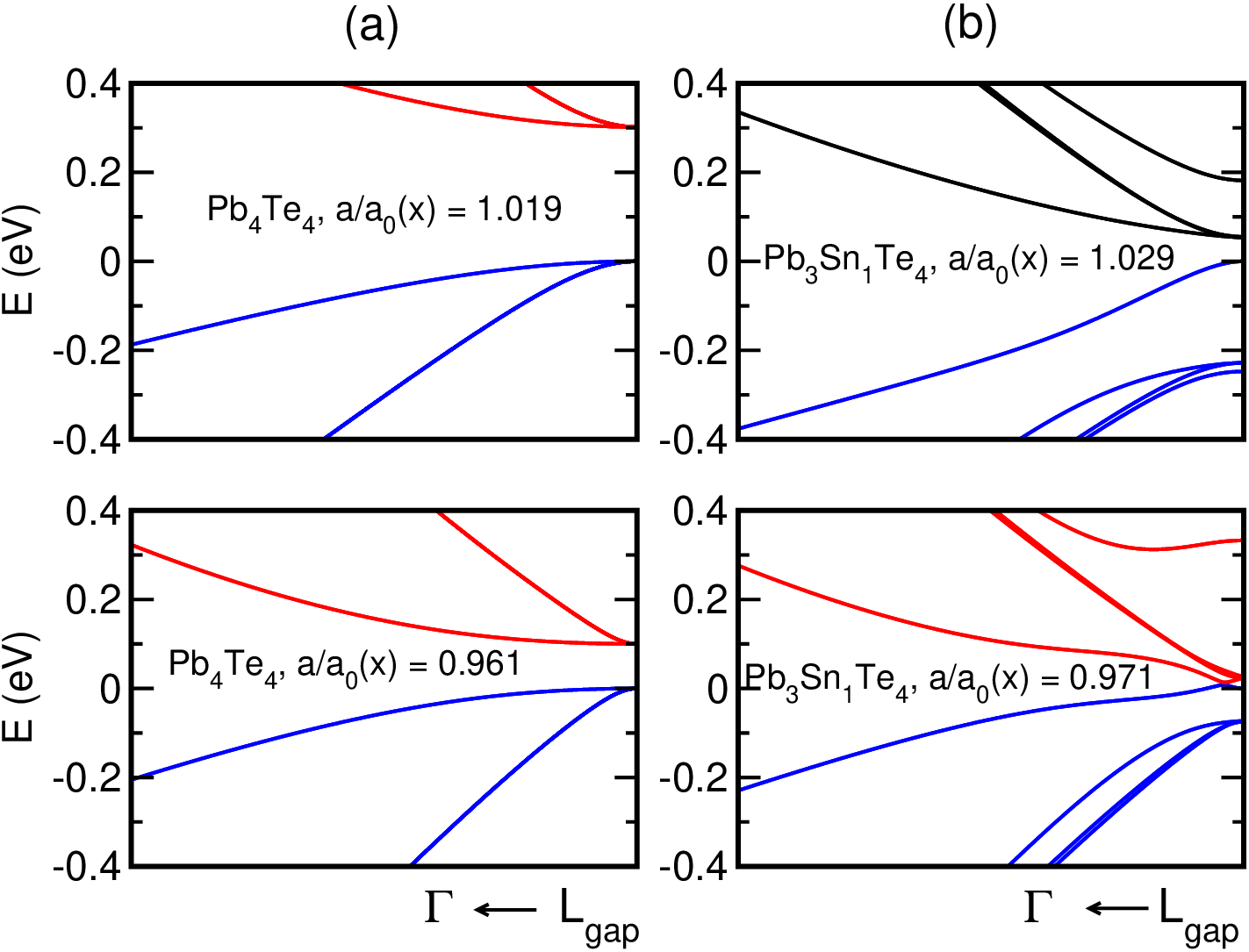}
\caption{\label{fig4} (color online) Band structures of 
Pb$_4$Te$_4$ (a) and 
of Pb$_3$Sn$_1$Te$_4$ (b) for topologically 
trivial (upper 
row) 
and nontrivial (lower row) cases. }
\end{figure}

In Fig. \ref{fig4} we compare the band 
structures of 
Pb$_4$Te$_4$ (left column) and of  
Pb$_3$Sn$_1$Te$_4$ (right 
column) for topologically trivial (upper 
row) and  nontrivial  (lower row) 
cases. Only the highest eight valence bands 
and the lowest eight 
conduction bands  are shown. The 
calculations are performed for lattice 
constants shown in the panels, where     
\begin{equation}  
a_0(x) = (6.46\ -0.16x)\ \angstrom  
\end{equation}  
is the equilibrium lattice constant 
according to the Vegard's law for \pst.   

The most important modification of the band 
structure relative to the VCA and caused by the distinction of 
the Pb and Sn atom is the  splitting of levels. 
We stress that this splitting 
is not related to a change of the  crystal 
point symmetry, which is $O_h$ 
for both PbTe and Pb$_3$Sn$_{1}$Te$_4$, but 
to the chemical difference between Pb and Sn.   

A closer inspection of the data in the 
lower right panel of the figure shows 
that for small lattice constants the main 
energy gap 
is shifted away from the $L_{gap}$ point, 
as in SnTe at equilibrium.\cite{nimtz,khoklov} 
When the lattice constant decreases this 
gap also closes, independently 
of the gap at the $L_{gap}$  point what leads to 
the changes in the topological 
characteristics of the band structure.  For 
Pb$_3$Sn$_{1}$Te$_4$  
the transition occurs for $a/a_0(x)\approx  
0.9897$ and 
for ${\bm k}\approx 0.482\times 
\frac{\pi}{a}(1,1,1)$.

The pressure dependencies of the band 
energies at $L_{gap}$ are shown in Fig. 
\ref{fig5} for the eight highest valence 
bands and the eight lowest 
conduction bands. The results for 
Pb$_3$Sn$_1$Te$_4$ (right panel) are 
compared with those for Pb$_4$Te$_4$ (left 
panel). 
These sixteen bands are occupied by eight 
electrons. 
In PbTe, the band inversion occurs at 
$a/a_0=0.97$, and for smaller $a$
electrons occupy cation rather than anion 
orbitals. 
In the case of Pb$_3$Sn$_1$Te$_4$ the 
situation is more complicated. 
As it follows from Fig.~\ref{fig5}, 
there is a large interval of $0.97 < 
a/a_0(x) < 1.02$, in which the band 
gap vanishes, i.e., the system is metallic, 
and electrons occupy 
combinations of the cation and anion $p$ 
orbitals. Finally, for $a < 
0.97$, the band gap is finite, the system 
is insulating, the 
band structure in inverted, and electrons 
occupy the cation orbitals only. 
In Fig.~\ref{fig6} we show energy gaps, 
contributions of $p$ 
orbitals of different atoms to the 
wave functions of the top valence  band  at 
$L_{gap}$, and the MCN for the 
valence bands as a function of 
the lattice constant. 
The value of $E_{gap}$ at $L_{gap}$ 
follows from the behavior of 
the energy levels, Fig.~\ref{fig5}. In 
addition to $C_{Te}$, we also show the 
contribution of cation $p$ orbitals to the 
wave function. Because the 
considered supercell has the $O_h$ point 
symmetry, it is possible to 
calculate the MCN. For $a/a_0 > 1.02$ 
MCN=0, and when 
$E_{gap}=0$ we  observe a jump to MCN=1. 
The interesting jump to 
MCN=-3 is related to the closing of the gap 
at a certain point along 
$\Gamma \rightarrow L_{gap}$ direction, 
discussed  above, what evidently leads 
to the change of topological properties of  
the  valence band structure. 
Finally, when $a/a_0 < 0.97$, then 
$E_{gap}$ is negative, the valence 
band at the $L_{gap}$ point is composed 
mainly from $p$(Sn) orbitals, and 
MCN=-2 like in the case of pure SnTe.

\begin{figure}
\end{figure}

\begin{figure}
\vspace{1cm}
\includegraphics[width=\linewidth]{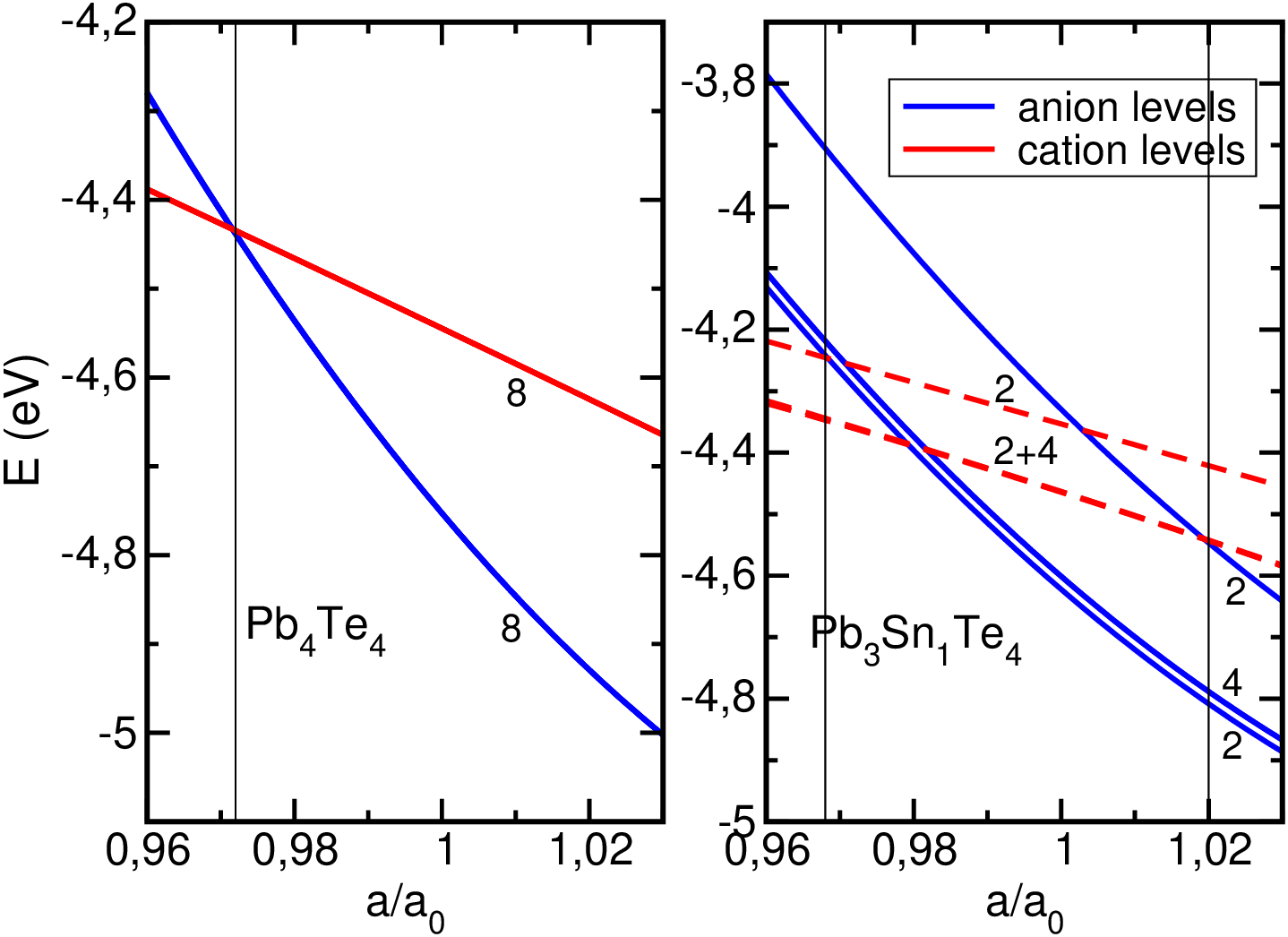}
\caption{\label{fig5} (color online)  Energy levels of the 
eight highest valence bands 
and the eight lowest conduction bands at 
the $L$ point of the Brillouin 
zone. 
The continuous and broken lines correspond 
to levels which 
wave functions are built mainly from $p$ 
orbitals of anions and cations, 
respectively. The numbers near the lines 
give the degeneracy of the 
levels (2+4 means that we have two lines, invisible 
in this scale, with the 
degeneracies 2 and 4, respectively). The thin 
vertical lines are the borders between 
topologically nontrivial, transition and 
topologically trivial regions of the 
lattice constant. Notice that for 
Pb$_4$Te$_4$ the width of the transition 
region is zero.}
\end{figure}
\begin{figure}
\end{figure}

\begin{figure}
\vspace{1cm}
\includegraphics[width=\linewidth]{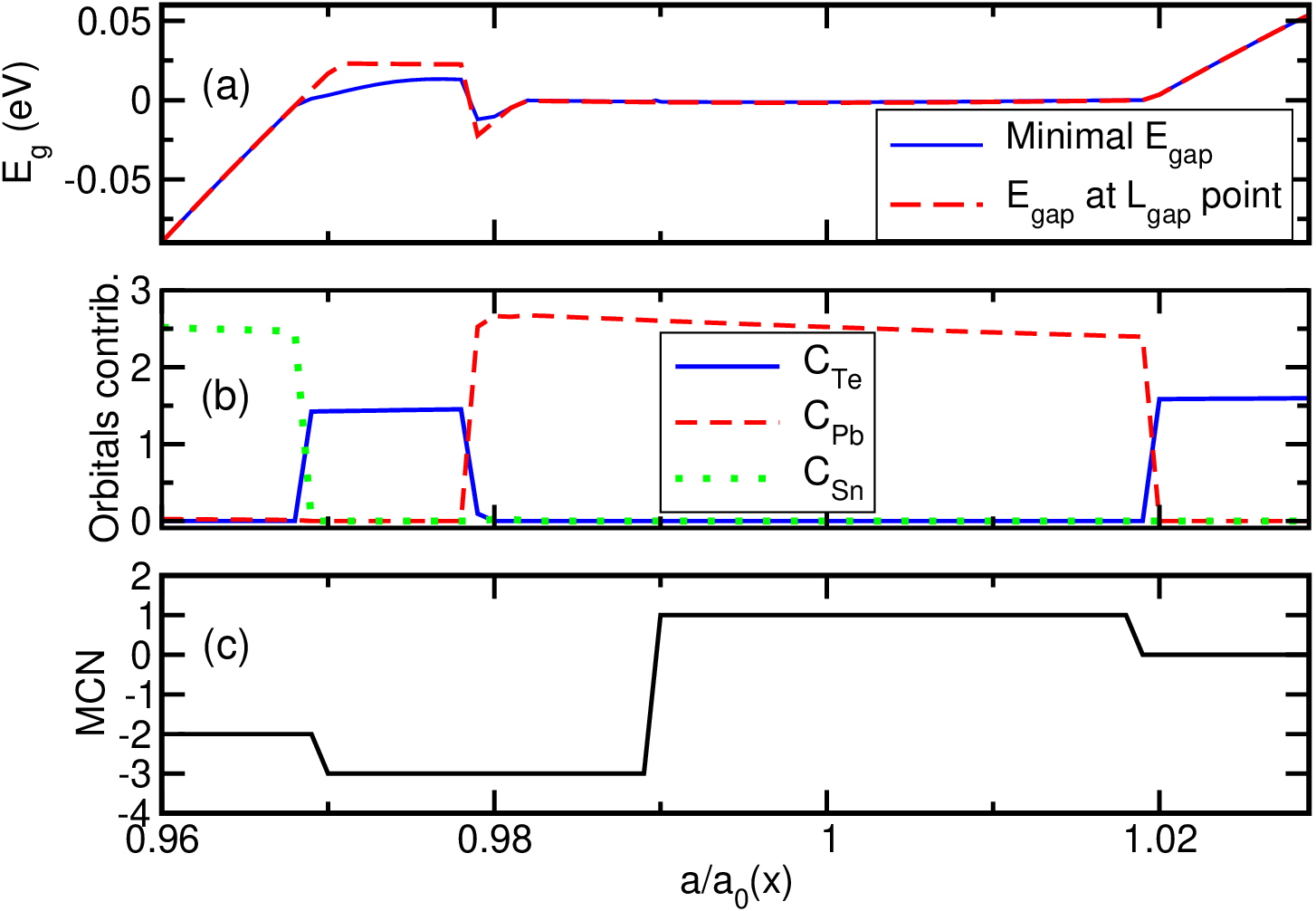}
\caption{\label{fig6} (color online)  
Band structure characteristics of 
Pb$_3$Sn$_1$Te$_4$. Dependence on the 
lattice constant of: 
(a) Minimal energy gap along $\Gamma 
\rightarrow\ L_{gap}$ direction and 
$E_{gap}$ at the $L_{gap}$, 
(b) Contributions of pseudoatomic $p$ 
orbitals to the valence band 
wave functions at $L_{gap}$, 
and (c) the Mirror Chern Number for valence 
bands. }
\end{figure}
\begin{figure}
\end{figure}

\begin{figure}
\includegraphics[width=\linewidth]{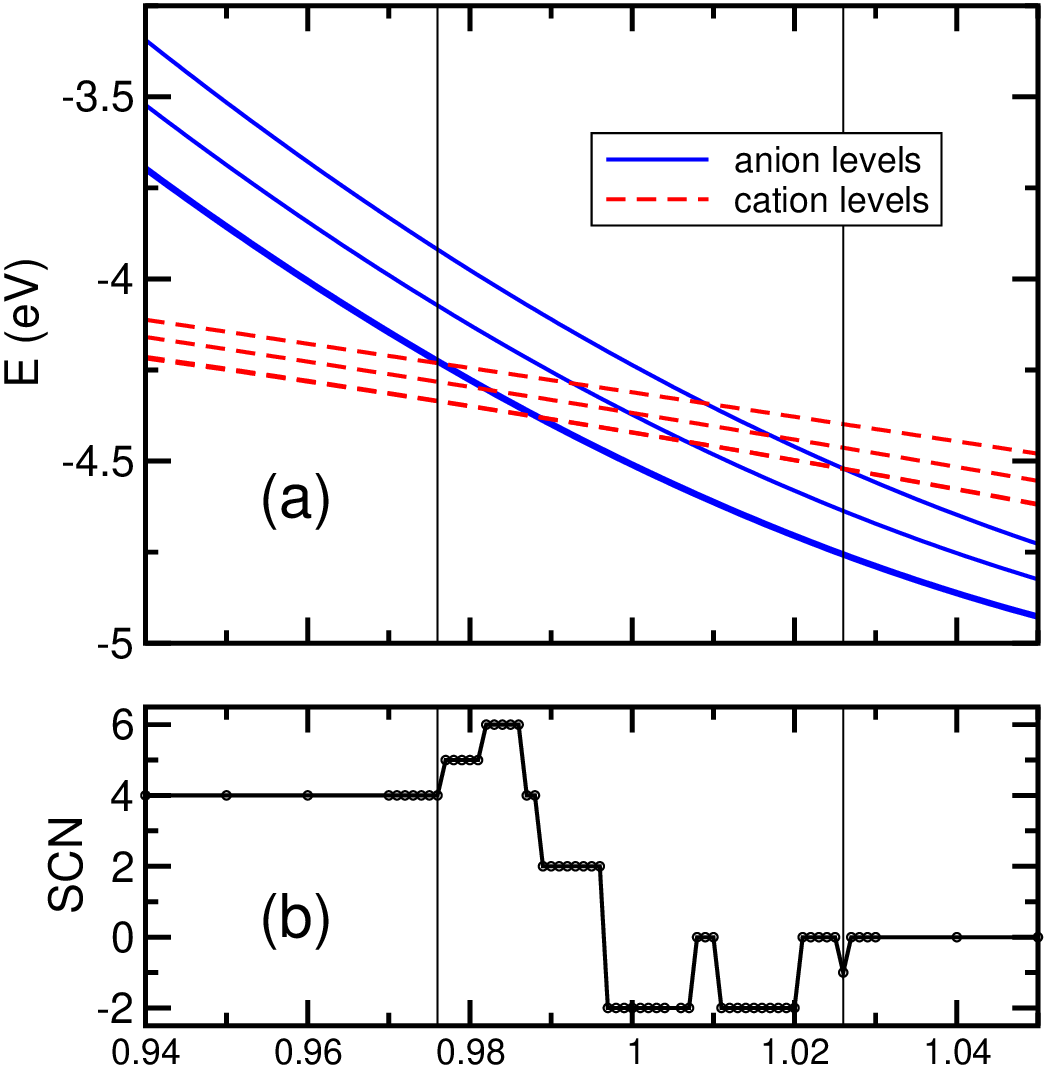}
\caption{\label{fig7}  (color online) Band  
energy levels at the $L_{gap}$ point for 
anions and cations a) and the Spin 
Chern Number b) as the functions of the 
lattice constant for one of the 
configurations for 16-atom supercell 
containing three Sn atoms. The thin 
vertical lines are the borders between 
topologically nontrivial, transition and 
topologically trivial regions of the 
lattice constant. }
\end{figure}

It turns out that this picture is very 
general. The behavior of  the energy levels 
shown in Fig.~\ref{fig5} is qualitatively 
very similar  for larger 
supercells with 16, 64, 216 atoms 
containing different  numbers of 
randomly distributed Sn atoms.  The main 
difference is that the 
degeneracy of levels is reduced not 
only  by the presence of two types of 
cations, but also by the reduction of the 
crystal point symmetry for 
typical atomic configurations.   We 
calculated the band structure 
characteristics for all nonequivalent 
atomic  configurations for the 16-atom 
$1\times 1\times 2$ supercell. (For larger 
supercells, the SCN were 
not calculated because they require 
non practically long computation 
times necessary to obtain convergent 
results.) As an example, in 
Fig.~\ref{fig7} we show the results for the 
$1\times 1\times 2$ 
supercell with three Sn atoms.  In the 
interval 0.97< $a/a_0(x)$<1.03 the 
energy gap is zero, and the anion  and 
cation energy levels are mixed up, 
what is reflected in the abrupt  variations 
of the band structure topology 
characterized by the SCN.  

The conclusions from calculations for 
$1\times 1\times 2$ supercell are 
as follows.

First, independent of the alloy composition 
(i.e., of number of cations in 
the supercell) and the spatial distribution 
of cations we always find a finite 
interval of the lattice constants in which 
$E_{gap}=0$. 

Second, when the energy gap closes we 
observe the abrupt drop in the 
values of $C_{Te}$ from $\sim$1.5  to 
almost zero.\\

Third, for a given plane 
in the BZ the SCNs or MCNs 
do not uniquely characterize the topology of the 
band structure. In most cases, 
when the gap closes with the decreasing 
lattice constant, we observe a 
jump in SCN from zero to a nonzero value. 
However this is not always 
the case. 
For example, for $1\times 1\times 2$ 
supercell containing two Sn atoms 
placed in such a way that they constitute a 
BCC lattice, the SCN does not 
change from the zero value when the gap 
closes. The same situation is 
found for 
MCN with respect to the (110) plane. 
Only for the MCN for the (001) 
plane, when the gap closes, we observe a 
jump to the value -2. 

Finally, 
for many of the considered configurations, 
we observe a number of jumps of SCN in the topologically 
nontrivial region of 
$a/a_0(x)$. This confirms that, what is 
obvious from the mathematical 
point of view, the topological 
characteristics based on Chern numbers do 
not uniquely characterize the valence band 
vector bundles. In general, the 
topological indices based on Chern numbers 
may distinguish two phases 
when the values of these 
numbers are different, but if the values 
are the same it is impossible to 
decide whether the studied phases are 
different or not. 

\begin{figure}
\end{figure}

\begin{figure}
\includegraphics[width=\linewidth]{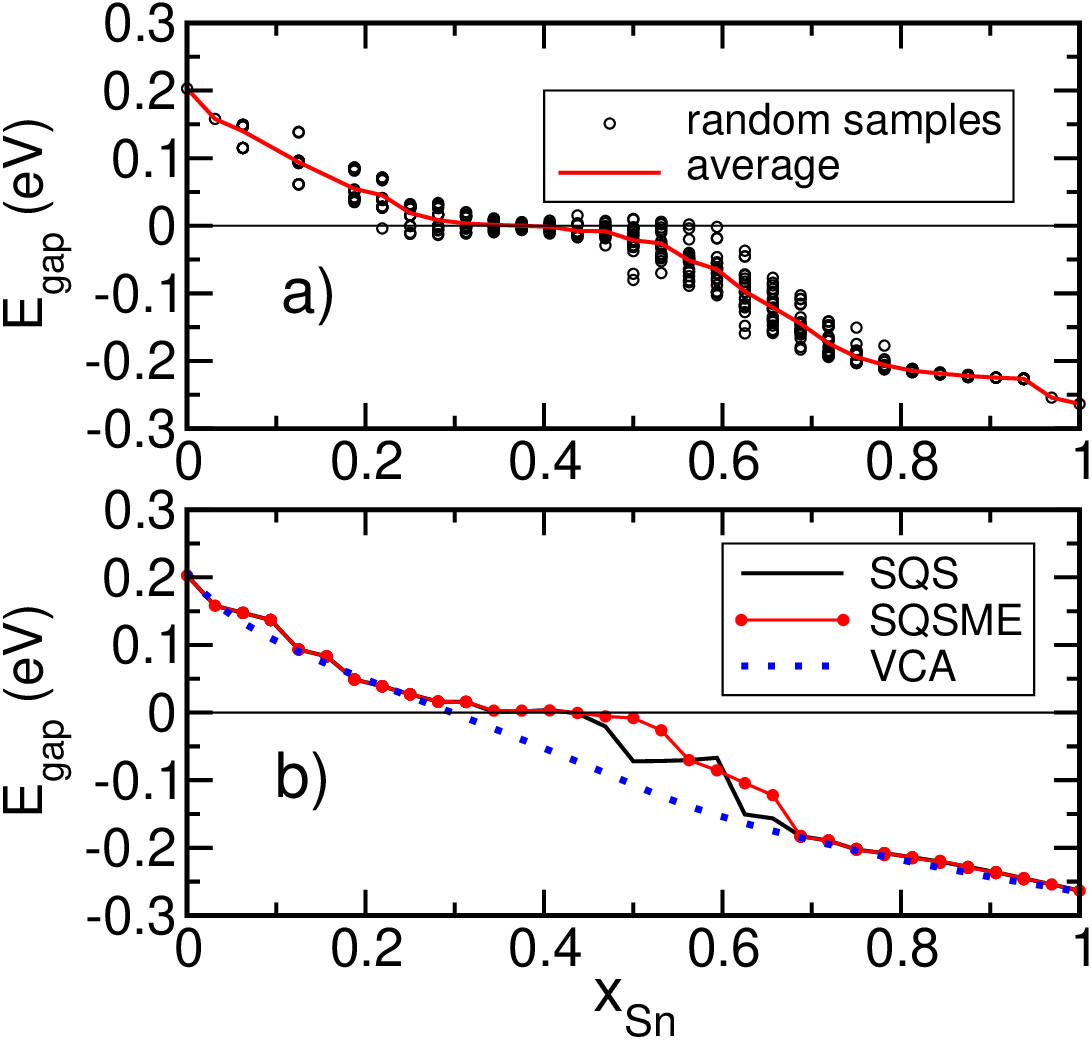}
\caption{\label{fig8}(color online) (a) 
Minimal energy gaps on [111] direction 
for 20 
randomly chosen tin configurations (points) 
and the average (solid 
line). (b) Minimal energy gaps on [111] 
direction for the best SQS tin 
configurations (solid line), for the 
configurations of minimal 
energies chosen from 10 best SQS (line with 
points) and for the VCA (dotted 
line).}
\end{figure}

\subsection{Composition dependence of \pst\ 
band gap}

The calculated composition dependence of 
the \pst\ band gap is shown in 
Fig.~\ref{fig8}. The results were obtained using three approaches, 
namely by averaging over 20 different random distributions of cations in 
the supercell, the Special Quasirandom Structures (SQS) 
method,\cite{zunger} and the VCA.

The $2\times 2\times 2$ supercells 
containing 64 atoms allow us to study 
energy gaps for a dense set of 32 
compositions. For each composition, 20 
random cation configurations were 
considered. In Fig.~\ref{fig8}a we 
show minimal energy gaps along the [111] 
direction. 
The first observation to make is that, for 
a given number of Sn/Pb atoms, the 
band gaps strongly depend on their 
distribution in the supercell. Indeed, the 
spread of $E_{gap}$ can be as high as 0.15 
eV, which is close to the band 
gap itself. However, for some composition 
ranges, especially for $x < 0.2$ 
and $x > 0.8$, several configurations give 
almost the same $E_{gap}$. (We 
do not have a convincing explanation of 
this effect.) 

Secondly, in spite of the large 
fluctuations, the gap averaged over the 
configurations, $E_{gap}^{ave}$, is a 
smooth function of composition. 
The composition dependence of 
$E_{gap}^{ave}$ is strongly 
non-linear, but the non-linearity does not 
consist in the typical parabolic bowing, 
which characterizes most of semiconductor 
alloys. 
The most prominent feature is that 
$E_{gap}^{ave}$ vanishes to within 0.01 eV 
in a wide composition range 
$0.3 < x < 0.45$. This effect can be 
explained based on the previous 
discussion of the energy levels splitting 
due to the presence of two different cations in the alloy.

Because our simple averaging procedure of 
$E_{gap}$ has no solid 
physical foundation, we calculated the band 
gap $E_{gap}^{SQS}$ for 
\pst\ using the SQS method. In this 
approach, one approximates a real random 
alloy without the translational symmetry by 
a periodic structure of the 
same composition chosen in such a way that 
certain parameter $\epsilon$, a measure describing the 
difference  between the atomic spatial 
distribution of this structure and of 
perfectly random system is minimized. In 
order to obtain best SQSs we proceeded 
along the 
prescription described in 
Ref.~\onlinecite{pezold} where the parameter 
$\epsilon$ is defined. The results for 
energy 
gaps obtained for our best SQSs for even 
number of tin atoms in the 
supercells are identical to those 
calculated using the Table 1 in 
Ref.~\onlinecite{pezold}. 

The SQS energy gaps $E_{gap}^{SQS}$  are 
shown in 
Fig.~\ref{fig8}b by the continuous line. The 
overall composition 
dependence of the $E_{gap}^{SQS}$ is quite 
close to the average 
$E_{gap}^{ave}$ shown in Fig.~\ref{fig8}a. 
In particular, 
$E_{gap}^{SQS}=0$ in the same composition 
interval $0.35 < x < 
0.45$. (We note that $E_{gap}^{SQS}$ is 
constant for $0.5 < x < 0.6$, 
but this seems to be an artifact of the 
method.) The SQS results of 
Fig.~\ref{fig8}b are qualitatively similar 
to those obtained in 
Ref.~\onlinecite{daw} where this problem 
was considered, although for 
smaller systems.

During the crystal growth there are two 
factors which decide about the 
placements of different cations in the 
crystal lattice. The first one is the configurational entropy, which 
promotes 
the alloy randomness, and the second 
factor is the total energy, which tends to assume a 
minimal value, which can result in alloy ordering or phase segregation. 
Although due to high 
temperature of the crystal growth and the 
finite time of the process it seems that 
the entropy factor is the most important one, however the second one 
should not be totally neglected. By  
construction, the SQS method takes into 
account only the entropy factor. In 
order to study the influence of the energy 
factor, instead of 
taking the best SQS configuration 
characterized by the smallest 
value of $\epsilon$, from about 300 000 configurations 
generated during the program's run we chose 
10 configurations 
characterized by smallest values of $\epsilon$. For a given tin's 
concentration the dispersion of $\epsilon$s 
for these 10 configurations was 
very small. The dispersion of energies 
depend on tin's concentration.  For 
$x < 0.3$ and for $x > 0.7$ it is 
negligible, however, for $0.3 < x < 
0.7$ the difference between the total 
energies for 
configurations with highest and lowest 
total energies is in the interval 0.001 
-- 0.12~eV. Taking for the band structure 
calculations the configuration 
of the lowest energy we obtain the SQSME 
(special quasirandom structure minimal 
energy) curve in Fig.~\ref{fig8}b. Although 
this curve is not perfectly smooth, 
the step-like behavior for 0.5 < x <0.6 of the SQS curve disappears.

For comparison, in Fig.~\ref{fig8}b we also 
show the predictions of VCA (dotted 
line).  
It is clear that the VCA fails in 
describing 
the band gap of \pst\ for a relatively wide 
range of compositions, in which the 
transition from the direct to the inverted 
band structure, or, in other words, 
from the topologically trivial to non-
trivial situation takes place. 

We see that SQS and SQSME results 
practically coincide for most compositions
except the small composition window $0.5 < x 
< 0.65$, where the smooth SQSME data seem 
to be more physical. Let us notice that the 
average band gap $E_{gap}^{ave}(x)$ 
shown in Fig.~\ref{fig8}a is also close to 
that obtained within the SQS/SQSME 
method. 
All these approaches explicitly 
differentiate between the two kinds of 
cations, Sn and Pb.
On the other hand, the VCA assumes an 
averaged cation, and neglects the effect of 
fluctuations.  

While the SQS method is a well justified 
procedure, our analysis of the gap for 
various configurations, Fig.~\ref{fig8}a, provide 
a deeper insight into the problem by showing the considerable 
impact of fluctuations of both composition 
and configurations. As a consequence, the 
spread in $E_{gap}$ shows that local composition 
fluctuations can result in efficient alloy 
scattering of 
carriers, which implies shortening of 
carrier lifetime.

\section{Conclusions} 

We analyzed topological properties of the mixed \pst\ crystals. Three 
approaches were used to calculate band structure, namely the virtual 
crystal approximation, the supercell method, and the special quasirandom 
structures method. The transition between the trivial and non-trivial 
topological phase can be driven either by 
the increasing content of Sn in \pst\, or by hydrostatic pressure for 
$x<0.3$, when the band gap at ambient pressure is positive. From our results 
the following 
conclusions may be drawn. 

1. For a compound crystalline semiconductors like PbTe, the transition from the 
trivial to the non-trivial 
topological phases is sharp and driven by the pressure-induced inversion 
of the band gap character. A sharp transition is predicted also for \pst\ 
mixed crystals at $x \approx 0.3$, but only when the VCA is used. 

2. Both the supercell method and the SQS approach explicitly 
differentiate between the two cations, Pb and Sn in \pst. This induces 
additional splittings of the energy bands, which are caused by different 
potentials of Pb and Sn and by the reduction of the crystal 
symmetry. In consequence, transitions from the trivial to the non-trivial 
phase are broad. The composition ranges in which the band structure is 
direct (0<x<0.3) and inverted (0.6<x<1) are separated by an 
unexpectedly large window 0.3<x<0.6 characterized by the vanishing 
band gap. Similarly, the pressure induced phase transition is broad, and 
extends over a window of 2 \%  of the lattice constant values, in which 
$E_{gap} \approx 0$. This shows that the sharpness of topological transitions 
predicted by the VCA is an artifact.  

3. In the regions of the compositions or of the lattice constants with the 
vanishing band gap, there is no criterion to decide whether the 
topological phase is trivial or nontrivial, because the mirror Chern 
number, the spin Chern number, and the orbital content of band edge 
wave functions vary rapidly and randomly. In this interval, topology of 
the band structure is very complicated, and sometimes the topological 
indices or the orbital content of the wave functions suggest triviality of 
the phase. To establish the phase, it is necessary to analyze the band 
structure in a larger interval of the lattice constants.  

4. Our results can explain the experimental studies of conductivity of 
\pst\ with $x=0.25$ reported recently in Ref. \onlinecite{liang}. The 
authors observe a pressure-driven transition from the insulating to the 
metallic phase at about 12 kbar, followed by the re-entrance of the 
insulating phase at 24 kbar. According to Fig.~\ref{fig8}, \pst\ with 
$x=0.25$ has a direct band structure with a positive band gap of about 
0.05 eV. 
Analysing the energy gap dependence on the lattice parameter for 64 atom
supercell containing 25\% of tin with SQS distribution of cations and
knowing that $dE_{gap}/dp$=70 meV/GPa \cite{nimtz, khoklov} it is possible to
predict that the band gap, with the increasing hydrostatic pressure, closes at 
about 2 kbar and and \pst\ is metallic up to 16 kbar.
 For higher pressures 
$E_{gap}<0$, and the alloy is insulating. Comparing those results with 
experiment we observe that the predicted transition pressure to the 
metallic phase is too low (which can be due to our inaccuracy in 
determining $E_{gap}$), but the range of pressures corresponding to the 
gapless situation, 14 kbar, fits well the experimental value, 12 kbar. 

5. Another experimental verification of our theoretical findings could be 
performed with low temperatures optical measurements in the THz range 
under hydrostatic pressure for crystals with low carrier concentrations. 
Although very challenging, such an experiment is feasible, as shown by 
pressure studies of infrared reflectivity in the closely related \pss\ TCI 
mixed crystals.\cite{xi} 



\begin{acknowledgments}
The authors acknowledges the 
support from NCN (Poland) 
research projects Nr. 
UMO-2016/23/B/ST3/03725 (A{\L}) and Nr. 
UMO-2014/15/B/ST3/03833 (TS). The suggestions 
and discussions with Professor 
Ryszard Buczko are kindly acknowledged. 
\end{acknowledgments}


\begin{thebibliography}{[1]}
\bibitem{nimtz}
G. Nimtz, and B. Schlicht, in: {\em Narrow-Gap Semiconductors} (ed. G. 
H\"ohler), 
Springer Tracts in Modern Physics, Vol. 98 (ed. G. Hohler)
(Springer-Verlag, Berlin 1983).
\bibitem{khoklov}
D. R. Khoklov (ed.) {\em Lead Chalcodenides: Physics and Applications} 
(Taylor 
and Francis, New York 2003). 
\bibitem{dziawa}
P. Dziawa, B. J. Kowalski, K. Dybko, R. Buczko, A. Szczerbakow, M. Szot, 
E. 
{\L}usakowska, T. Balasubramanian, B. M. Wojek, M. H. Berntsen, O. 
Tjernberg, 
and T. Story, Nature Mat. {\bf 
11}, 1023 (2012).
\bibitem{tanaka}
Y. Tanaka, Z. Ren, T. Sato, K. Nakayama, S. Souma, T. Takahashi, K. 
Segawa, and 
 Y. Ando, Nat. Phys. {\bf 8}, 800 (2012)
\bibitem{xu}
S.-Y. Xu, C. Liu, N. Alidoust, M. Neupane, D. Qian, I. Belopolski, J.D. 
Denlinger, Y.J. Wang, H. Lin, L.A. Wray, G. Landolt, B. Slomski, J.H. 
Dil, A. 
Marcinkova, E. Morosan, Q. Gibson, R. Sankar, F.C. Chou, R.J. Cava, A. 
Bansil, 
and M.Z. Hasan, Nat. Commun. {\bf 3}, 1192 (2012)
\bibitem{wojek}
B. M. Wojek, P. Dziawa, B. J. Kowalski, A. Szczerbakow, A. M. Black-
Schaffer, M. 
H. Berntsen, T. Balasubramanian, T. Story, and O. Tjernberg, Phys. Rev. B 
{\bf 90}, 161202(R) (2014)
\bibitem{barone}
P. Barone, T. Rauch, D. Di Sante, J. Henk, I. Mertig, and 
S. Picozzi, Phys. Rev. B {\bf 88}, 045207 (2013)
\bibitem{rb1}
B. M. Wojek, R. Buczko, S. Safaei, P. Dziawa, B. J. Kowalski, M. H. 
Berntsen,
T. Balasubramanian, M. Leandersson, A. Szczerbakow, P. Kacman, T. Story,
and O. Tjernberg, Phys. Rev. B {\bf 87}, 115106 (2013)
\bibitem{rb2}
S. Safaei, P. Kacman, and R. Buczko, Phys. Rev. B {\bf 88}, 045305 (2013)
\bibitem{rb3}
S. Safaei, M. Galicka, P. Kacman, and R. Buczko, New J. Phys. {\bf 17}, 
063041 
(2015)
\bibitem{hsieh}
T. H. Hsieh, H. Lin, J. Liu, W. Duan, A. 
Bansil, and L. Fu, Nature Comm. 
{\bf 
3}, 982 (2012).
\bibitem{fu_kane}
L. Fu, and C. L. Kane, Phys. Rev. Lett. {\bf 109}, 246605 (2012)
\bibitem{prodan}
E. Prodan, {\em Phys. Rev. B} {\bf 80}, 
125327 (2009)
\bibitem{lusak_jasz}
A. Lusakowski 46$^{th}$ International School \& Conference on the Physics 
of 
Semiconductors, "Jaszowiec 2017", unpublished
\bibitem{zunger}
A. Zunger, S.-H. Wei, L. G. Ferreira, and 
J. E. Bernard, Phys. Rev. Lett. {\bf 65}, 353 (1990)

\bibitem{openmx}
see http://www.openmx-square.org
\bibitem{CA}
D. M. Ceperley, and B. J. Alder, Phys. Rev. 
Lett. \textbf{45}, 566
(1980).
\bibitem{lusakowski1}
A. {\L}usakowski, P. Bogus{\l}awski, and T. 
Radzy\'nski, Phys.~Rev.~B 
{\bf 83}, 
115206 (2011)
\bibitem{suzuki}
T. Fukui, Y. Hatsugai, and H. Suzuki, {\em J. 
Phys. Soc. Jpn.} {\bf 74}, 1674 
(2005)
\bibitem{pezold}
J. von Pezold, A. Dick, M. Fri\'ak, and J. 
Neugebauer, Phys. Rev. B {\bf 
81}, 
094203 (2010)
\bibitem{daw}
X. Gao, and M.~S.~Daw, Phys. Rev. B {\bf 77}, 033103 (2008)
\bibitem{liang}
T. Liang, S. Kushwaha, J. Kim, Q. Gibson, J. Lin, N. Kioussis, R. J. 
Cava, and 
N. P. Ong, Sci. Adv. {\bf 3}, e1602510 (2017) 
\bibitem{xi}
X. Xi, X.-G. He, F. Guan, Z. Liu, R. D. Zhong, J. A. Schneeloch, T. S. 
Liu, G. 
D. Gu, X. Du, Z. Chen, X. G. Hong, W. Ku, and G. L. Carr, Phys. Rev. 
Lett. {\bf 113}, 
096401 (2014).

\end{thebibliography}
\end{document}